\documentclass{llncs}
\newfont{\mycrnotice}{ptmr8t at 7pt}
\newfont{\myconfname}{ptmri8t at 7pt}
\usepackage[utf8]{inputenc}
\usepackage[T1]{fontenc}
\usepackage{mathptmx}
\usepackage{amssymb}
\usepackage{caption}
\usepackage{graphicx}
\usepackage{epstopdf}
\usepackage{epsfig,graphicx}
\usepackage{color}
\usepackage{url}
\usepackage{tabularx}
\usepackage{xfrac}
\usepackage{array}
\usepackage[labelfont=bf]{caption}
\usepackage{color}
\usepackage{subcaption}
\usepackage{multirow}
\usepackage{enumitem}
\usepackage{amsmath}
\usepackage{algorithm,algorithmic}
\usepackage{hhline}
 \usepackage{tikz}
\usepackage{booktabs}
\usepackage{adjustbox}
\let\subparagraph\relax
\usepackage{titlesec}
\titlespacing\section{1pt}{12pt plus 2pt minus 1pt}{0pt plus 1pt minus 1pt}
\titlespacing\subsection{1pt}{12pt plus 2pt minus 1pt}{0pt plus 1pt minus 1pt}
\titlespacing\subsubsection{1pt}{12pt plus 2pt minus 1pt}{0pt plus 1pt minus 1pt}
\usepackage{hyperref}
\definecolor{darkblue}{rgb}{0,0,.7}
\hypersetup{
     colorlinks   = true,
     citecolor    = blue,
     urlcolor     = blue,
     linkcolor    = blue
     }

\begin{document}


\title{Multiple Models for Recommending Temporal Aspects of Entities}

\author{
Tu Nguyen\inst{1}, Nattiya Kanhabua\inst{2}, Wolfgang Nejdl \inst{1}\\
 \institute{L3S Research Center / Leibniz Universit\"{a}t Hannover, Hannover, Germany \\
  \email{\{tunguyen,nejdl\}@L3S.de}
  \and 
  NTENT España, Barcelona, Spain\\
\email{nkanhabua@NTENT.com}
}}
\maketitle
\vspace{-0.2cm}
\begin{abstract}
Entity aspect recommendation is an emerging task in semantic search
that helps users discover serendipitous and prominent information with
respect to an entity, of which \textit{salience} (e.g., popularity) is
the most important factor in previous work. However, entity aspects
are temporally dynamic and often driven by events happening over
time. For such cases, aspect suggestion based solely on salience features can
give unsatisfactory results, for two reasons. First, salience is often
accumulated over a long time period and does not account for
\textit{recency}. Second, many aspects related to an event entity are
strongly time-dependent. In this paper, we study the task of temporal
aspect recommendation for a given entity, which aims at recommending
the most relevant aspects and takes into account time in order to
improve search experience. We propose a novel \textit{event-centric}
ensemble ranking method that learns from multiple time and
type-dependent models and dynamically trades off salience and recency
characteristics. Through extensive experiments on real-world query
logs, we demonstrate that our method is robust and achieves better
effectiveness than competitive baselines.

\end{abstract}


%

\vspace{-0.2cm}
\vspace{-0.2cm}
\section{Introduction}
\label{sec:introduction}



Beyond the traditional ``ten blue links'', to enhance user experience
with entity-aware intents, search engines have started including more
semantic information, i.e., (1) suggesting related
entities~\cite{blanco2013entity,fischer2015timely,yu2014building,zhang2016probabilistic}, and
(2) supporting entity-oriented query completion or complex search with
additional information or 
\textit{aspects}~\cite{balog2016report,reinanda2015mining,spina2012identifying}. These
aspects cover a wide range of issues and include (but are not limited
to) types, attributes/properties, relationships or other entities in
general. They can also be influenced by changes over time, as the focus of \textit{public attention} shifts between different aspects. To improve the recommendation of these entity aspects, it is essential to consider this dynamic nature over the temporal dimension.

Exploiting collaborative knowledge bases such as Wikipedia and
Freebase is a common practice in semantic search, i.e., by exploiting anchor
texts and inter-entity links, category structure, internal link
structure or entity types~\cite{blanco2013entity}. More recently,
researchers have also started to integrate knowledge bases with query
logs for \textit{temporal} entity knowledge
mining~\cite{chirigati2016knowledge,yu2014building}. In this work, we
address \textit{the temporal dynamics of recommending entity aspects}
and also utilize query logs, for two reasons. First, query logs are
strongly entity related: more than 70\% of Web search queries contain
entity information ~\cite{lin2012active,pound2010ad}. Queries often
also contain a short and very specific piece of text that represents
users' intents, making it an ideal source for mining entity aspects.
Second, different from knowledge-bases, query logs naturally capture
temporal dynamics around entities. The intent of entity-centric
queries is often triggered by a current
event~\cite{Kulkarni:2011:UTQ:1935826.1935862,kong2015predicting}, or
is related to ``what is happening right now''.

\begin{figure}[t]
\centering
		\includegraphics[width=0.7\columnwidth]{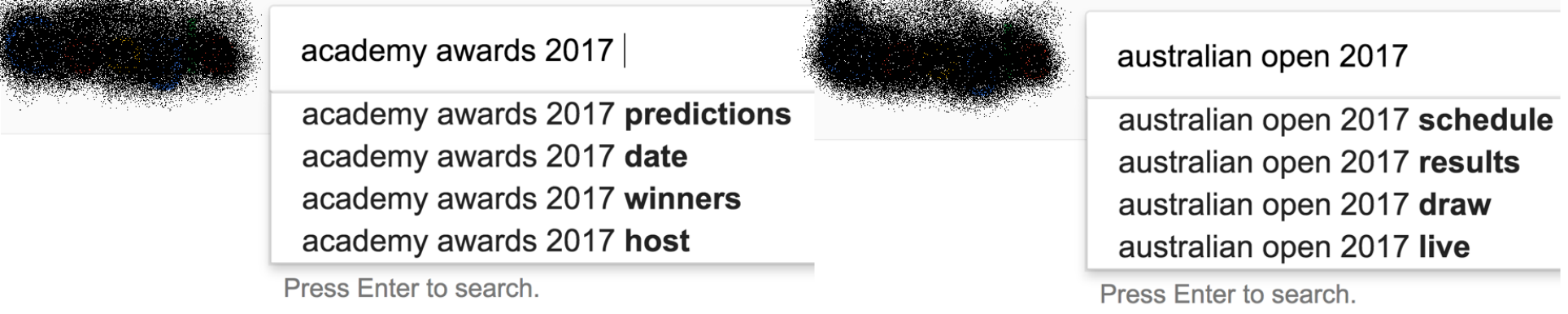}\vspace{-0.1cm}
		\caption{[Screenshot] Recommendation generated by a
		commercial search engine for \textsf{academy awards
		2017} and \textsf{australia open 2017}, submitted on
		March $31^{th}$, 2017, on a clean history browser.} 
		\label{fig:showcase0}
		\vspace*{-5pt}
\end{figure}

Previous work does not address the problem of temporal aspect
recommendation for entities, which are often event-driven. The task requires
taking into account the impact of temporal aspect dynamics and
explicitly considering the relevance of an aspect with respect to the
time period of a related event. To demonstrate the characteristics of
these entity aspects, we showcase a real search scenario, where entity
aspects are suggested in the form of query suggestion /
auto-completion, given the entity name as a
prior. Figure~\ref{fig:showcase0} shows the lists of aspect
suggestions generated by a well-known commercial search engine
for \textsf{academy awards 2017} and \textsf{australia open
2017}. These suggestions indicate that the top-ranked aspects are
mostly time-sensitive, and as the two events had just ended, the
recommended aspects are timeliness-wise irrelevant
(e.g., \textit{live}, \textit{predictions}).

\begin{figure*}[t!]
\centering
   \begin{subfigure}[t]{0.48\textwidth}
   \includegraphics[width=1\linewidth]{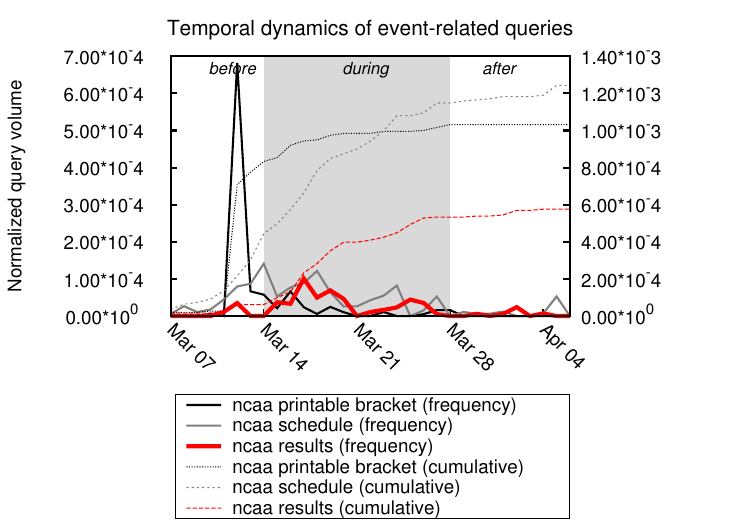}
   \caption{AOL}
   \label{fig:showcase2} 
\end{subfigure}
\begin{subfigure}[t]{0.48\textwidth}
   \includegraphics[width=1\linewidth]{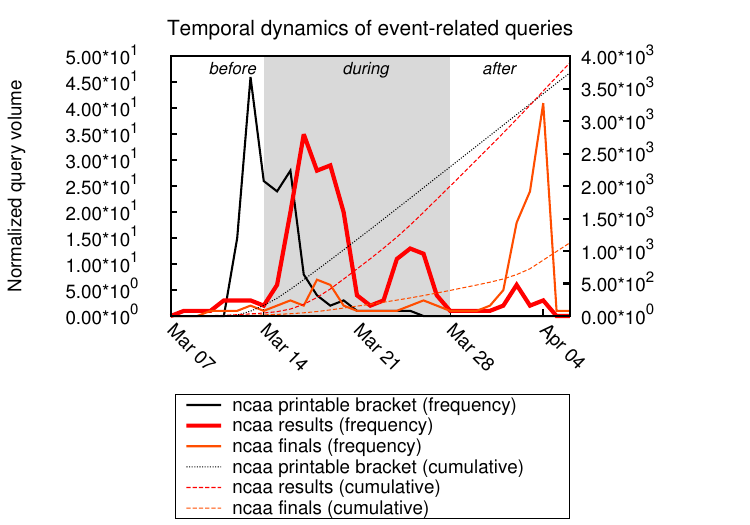}
   \caption{Google}
   \label{fig:Ng2}
\end{subfigure}
\caption{Dynamic aspect behaviors for entity \textsf{ncaa} in AOL and Google.}
\label{fig:showcase1}
\vspace*{-5pt}
\end{figure*}

While the precise methods employed by the search engine for its recommendations remain undisclosed, the subpar performance could potentially be attributed by the influence of aspect salience (in this case, query popularity) and the occurrence of the \textit{rich get richer} phenomenon: the salience of an aspect is
accumulated over a long time period. Figure~\ref{fig:showcase1}
illustrates changes in popularity of relevant searches captured in the
AOL (left) and Google (right) query logs (e.g., \textsf{ncaa printable bracket}, \textsf{ncaa
schedule}, and \textsf{ncaa finals}) for the NCAA\footnote{A
major sports competition in the US held annually by the National
Collegiate Athletic Association
(NCAA)- \url{https://en.wikipedia.org/wiki/Ncaa}.} tournament. The
basketball event began on March 14, 2006, and concluded on April 3,
2006. In order to better understand this issue, we present two types
of popularity changes, namely, (1)~frequency or query volume
(aggregated daily), and cumulative frequency.  Frequencies of
pre-event activities like \textsf{printable bracket}
and \textsf{schedule} gain increased volume over time, especially in
the \textit{before} event period. On the other hand, up-to-date
information about the event, such as, \textsf{ncaa results} rises in
importance when the event has started (on March 14), with very low
query volume before the event. While the popularity
of \textsf{results} or \textsf{finals} aspect exceeds that
of \textsf{ncaa printable bracket} significantly in the
periods \textit{during} and \textit{after} event, the cumulative
frequency of the pre-event aspect stays high. We witness similar phenomenon with the same event in 2017 in the Google query logs. We therefore postulate that (1)~long-term salience should provide good ranking results for the
periods \textit{before} and \textit{during}, whereas (2)~short-term or
recent interest should be favored on triggers or when the temporal
characteristics of an event entity change, e.g.,
from \textit{before}/\textit{during} to \textit{after} phase.  Different
event types (\textsf{breaking} or \textsf{anticipated} events) may
vary significantly in term of the impact of events, which entails
different treatments with respect to a ranking model.

Our contributions can be summarized as follows.
\vspace{-0.1cm}
\begin{itemize}
	\item We present the first study of temporal entity aspect
	recommendation that explicitly models triggered event time and
	type.
	\item We propose a learning method to identify time period
	and event type using a set of features that capture
	temporal dynamics related to event diffusion.  
	\item We propose a novel event-centric ensemble ranking method
	that relies on multiple time and type-specific models for
	different event entities. 
\end{itemize}
\vspace{-0.2cm}
To this end, we evaluated our proposed approach through experiments using real-world web search logs -- in conjunction with Wikipedia as
background-knowledge repository.



\vspace{-0.2cm}	
\section{Related Work}
\label{sec:relatedwork}

Entity aspect identification has been studied
in~\cite{spina2012identifying,reinanda2015mining}. \cite{spina2012identifying}
focuses on salient ranking features in microblogs. Reinanda et
al. ~\cite{reinanda2015mining} start from the task of mining entity
aspects in the query logs, then propose salience-favor methods for
ranking and recommending these aspects. When regarding an aspect as an
entity, related work connected to temporal IR is
~\cite{zhang2016probabilistic}, where they study the task of
time-aware entity recommendation using a probabilistic approach. The
method also \textit{implicitly} considers event times as triggering
sources of temporal dynamics, yet relies on coarse-grained (monthly)
granularity and does not recognize different phases of the event. It
is therefore not really suitable for recommending fine-grained,
temporal aspects. `Static' entity recommendation was first introduced
by the Spark~\cite{blanco2013entity} system developed at Yahoo!. They
extract several features from a variety of data sources and use a
machine learning model to recommend entities to a Web search
query. Following Spark, Sundog~\cite{fischer2015timely} aims to
improve entity recommendation, in particular with respect to
freshness, by exploiting Web search log data.  The system uses a
stream processing based implementation. In addition, Yu et
al.~\cite{yu2014building} leverage user click logs and entity pane
logs for global and personalized entity recommendation. These methods
are tailored to ranking entities, and face the same problems
as~\cite{zhang2016probabilistic} when trying to generalize to
`aspects'.

It is also possible to relate these entity aspects to RDF properties /
relations in knowledge bases such as FreeBase or
Yago. ~\cite{vadrevu2016ranking,dessi2016machine} propose solutions
for ranking these properties based on salience. Hasibi et
al.~\cite{hasibi2017dynamic} introduce dynamic fact based ranking
(property-object pairs towards a sourced entity), also based
on \textit{importance} and \textit{relevance}.  These properties from
traditional Knowledge Bases are often too specific (fact-centric) and
temporally static.

\section{Background and Problem Statement}

\subsection{Preliminaries}
\label{subsec:preliminaries}
In this work, we leverage clues from entity-bearing queries. Hence, we
first revisit the well-established notions of query logs and query-flow
graphs. Then, we introduce necessary terminologies and concepts for
entities and aspects. We will employ user log data in the form of
queries and clicks.

Our datasets consist of a set of queries $Q$, a set of URLs $U$ and
click-through information $S$. Each query $q \in Q$ contains query
terms $\textit{term}(q)$, timestamps of queries $\textit{time}(q)$
(so-called \textit{hitting time}), and an anonymized ID of the user
submitted the query.
A clicked URL $u \in U_q$ refers to a Web document returned as an
answer for a given query $q$.  Click-through information is a
transactional record per query for each URL clicked, i.e., an
associated query $q$, a clicked URL $u$, the position on result page,
and its timestamps. A co-clicked query-URL graph is a bipartite graph
$G = (V,E)$ with two types of nodes: query nodes $V_Q$ and URL nodes
$V_U$, such that $V = V_Q \cup V_U$ and $E \subseteq V_Q \times V_U$.
\vspace{-5pt}
\subsection{Problem Definitions}
\label{subsec:problem}

We will approach the task of recommending temporal entity aspect as a
ranking task. We first define the notions of an \textit{entity query},
a \textit{temporal entity aspect}, developed from the definition of
entity aspect in~\cite{reinanda2015mining}, and an \textit{event
  entity }. We then formulate the task of recommending temporal entity
aspects.

\textbf{Definition 1}. \textit{An entity query $q_{e}$ is a query that
  is represented by one Wikipedia entity $e$}. We consider $q_{e}$ as
the representation of $e$.

\textbf{Definition 2}. \textit{Given a ``search task'' defined as an
  atomic information need, a temporal ``entity aspect'' is an
  entity-oriented search task with time-aware intent.}. An
entity-oriented search task is a set of queries that represent a
common task in the context of an entity, grouped together if they have
the same intent~\cite{reinanda2015mining}. We will use the notion of
query $q$ to indicate an entity aspect $a$ interchangeably hereafter.

\textbf{Definition 3}. \textit{An entity that is related to a near
  event at time $t_{i}$ is called an event-related entity, or event
  entity for short.} Relatedness is indicated by the observation that
\textit{public attention} of temporal entity aspects is triggered by
the event. We can generalize the term \textit{event entity} to
represent any entity that is related to or influenced by the event. An
event entity $e$ that is associated to the event whose type
$\mathcal{C}$ can be either \textit{breaking} or \textit{anticipated}.
An event entity is also represented as a query with hitting time
t. The association between t and the event time --defines $e$'s time
period $\mathcal{T}$-- that can be either of the \textit{before},
\textit{during} or \textit{after} phases of the event. When the entity
is no longer event-related, it is considered a ``static'' entity.

\textbf{Problem (Temporal Entity-Aspect Recommendation):}
\textit{Given an event entity $e$ and hitting time $t$ as input, find
  the ranked list of entity aspects that most relevant with regards to
  e and t.}

Different from time-aware entity
recommendation~\cite{zhang2016probabilistic,tran2017beyond}, for an
entity query with exploratory intent, users are not just interested in
related entities, but also entity aspects (which can be a topic, a
concept or even an entity); these provide more complete and useful
information. These aspects are very time-sensitive especially when the
original entity is about an event. In this work, we use the notion of
\textit{event entity}, which is generalized to indicated related
entities of any trending events. For example, Moonlight and Emma Stone
are related entities for the 89th Academy Awards event. We will handle
the aspects for such entities in a temporally aware manner.


\vspace{-0.2cm}
\section{Our Approach}
\label{sec:ranking}

\begin{figure*}[t]
\centering
		\includegraphics[width=0.8\textwidth]{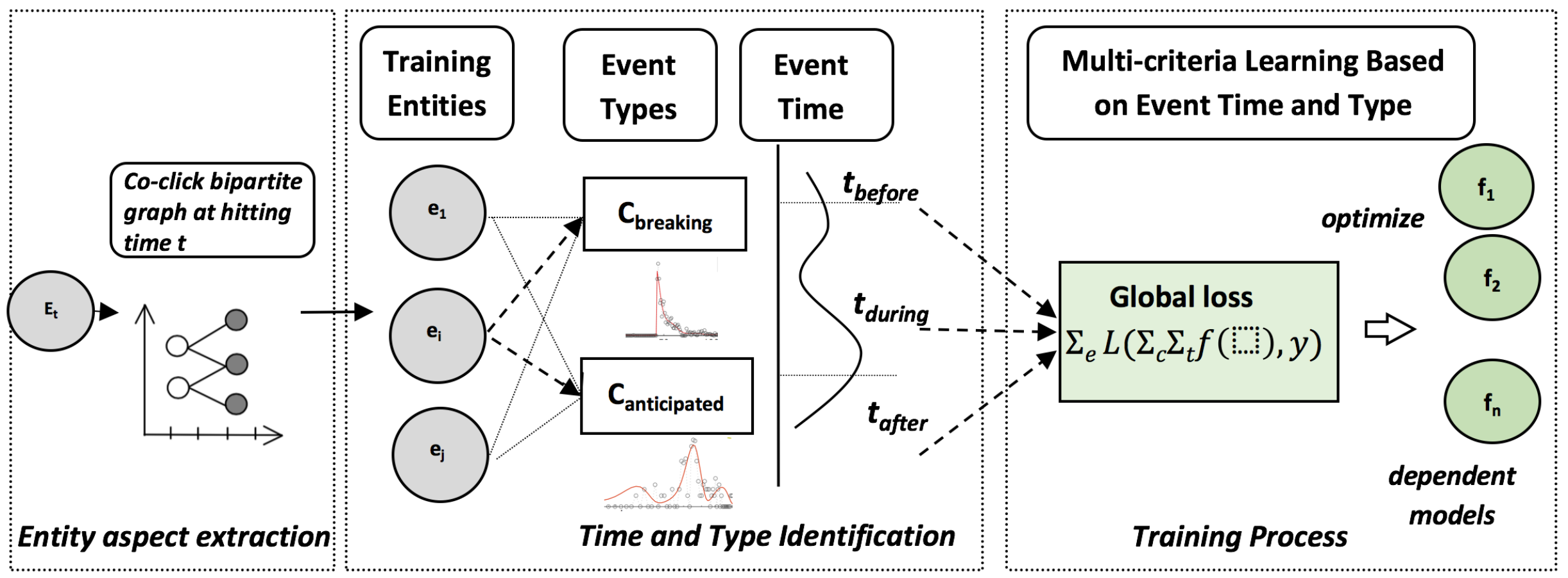}\vspace{-0.1cm}
		\caption{Learning time and type-specific ranking models.}\label{fig:multiplelearning}
\vspace{-0.1cm}
\end{figure*}

As event entity identification has been well-explored in related
work~\cite{kanhabua2015learning,kanhabua2016learning,KarmakerSantu:2017:MIP:3041021.3054188},
we do not suggest a specific method, and just assume the use of an
appropriate method. Given an event entity, we then apply our aspect
recommendation method, which is composed of three main steps. We
summarize the general idea of our approach in
Figure~\ref{fig:multiplelearning}. First, we extract suggestion
candidates using a bipartite graph of co-clicked query-URLs generated
at hitting time. After the aspect extraction, we propose
a \textit{two-step} unified framework for our entity aspect ranking
problem. The first step is to identify event type and time in a joint
learning approach. Based on that, in the second step, we divide the
training task to different sub-tasks that correspond to specific event
type and time. Our intuition here is that the timeliness (or
short-term interest) feature-group might work better for specific
subsets such as breaking and after events and vice versa. Dividing the
training will avoid timeliness and salience competing with each other
and maximize their effectiveness. However, identifying time and type
of an event on-the-fly is not a trivial task, and breaking the
training data into smaller parts limits the learning power of the
individual models. We therefore opt for an ensemble approach that can
utilize the whole training data to (1) supplement the uncertainties of
the time-and-type classification in the first step and (2) leverage
the learning power of the sub-models in step 2. In the rest of this
section, we explain our proposed approach in more detail.


\subsection{Aspect Extraction}
\label{subsec:graph}

The main idea of our approach for extracting aspects is to find
related entity-bearing queries; then group them into different
clusters, based on \textit{lexical} and \textit{semantic} similarity,
such that each cluster represents a distinct aspect. The
click-through information can help identifying related
queries~\cite{DBLP:journals/ftir/Silvestri10} by exploiting the
assumption that any two queries which share many clicked URLs are
likely to be related to each other.

For a given entity query $e$, we perform the following steps to find
aspect candidates. We retrieve a set of URLs $U_{e}$ that were clicked
for $e$ from the beginning of query logs until the hitting time
$t_{e}$. For each $u_j \in U_{e}$, we find a set of distinct queries
for which $u_j$ has been clicked. We give a weight $w$ to each
query-URL by normalizing \textit{click frequency} and \textit{inverse
query frequency} (CF-IQF)~\cite{Deng:2009:EMQ:1571941.1572001}, which
calculate the importance of a click, based on click frequency and
inverse query frequency. $CF-IQF$ = $cf \cdot log(N/(qf+1))$, where
$N$ is the number of distinct queries. A high weight $CF-IQF$
indicates a high click frequency for the query-URL pair and a low
query frequency associated with the URL in the whole query log. To
extract aspect candidates from the click bipartite graph, we employ a
personalized random walk to consider only one side of the query
vertices of the graph (we denote this approach as \textbf{RWR}). This
results in a set of related queries (aspects) to the source entity
$e$, ranked by click-flow relatedness score. To this end, we refine
these extracted aspects by clustering them using Affinity Propagation (AP)
on the similarity matrix of \textit{lexical} and \textit{semantic}
similarities. For semantic measure, we use a \textit{word2vec}
skip-gram model trained with the English Wikipedia corpus from the
same time as the query logs.  We pick one aspect with highest
frequency to represent each cluster, then select top-k aspects by
ranking them using RWR relatedness scores~\footnote{About complexity analysis, the  click bipartite graph construction costs $O(m+n)$ and RWR in practice, can be bounded by $O(m+n)$ for top-k proximity nodes. Note that $m,n$ are the number of edges and nodes respectively. AP is quadratic $O(kn^{2})$ time, (with k is the number of iterations), of our choice as we aim for a simple and effective algorithm and our aspect candidate sets are not large. A more efficient algorithm such as the Hierarchical AP can be used when candidate sets are large. The cost of constructing the similarity matrix is $O(n^{2})$.}. 



\subsection{Time and Type Identification}
\label{subsec:timeandtopic} 

Our goal is to identify the probability that an event-related entity
is of a specific event type, and in what time period of the event. We
define these two targets as a joint-learning time-series
classification task, that is based on event diffusion. In the
following, we first present the feature set for the joint-learning
task, then explain the learning model. Last we propose a light-weight
clustering approach that leverages the learning features, to integrate
with the ranking model in Section~\ref{subsec:ensemble}.


\textit{\textbf{Features.}} We propose a set of time series features
for our multi-class classification task. \textit{seasonality}
and \textit{periodicity} are good features to capture
the \textit{anticipated} -recurrent events. In addition, we use
additional features to model the temporal dynamics of the entity at
studied/hitting time $t_{e}$. We leverage query logs and Wikipedia
revision edits as the data sources for \textit{short}
and \textit{long} span time series construction, denoted as
$\mathcal{\psi}^{(e)}_{Q}$ and $\mathcal{\psi}^{(e)}_{WE}$ (for
seasonal, periodical event signals) respectively~\footnote{Wikipedia
page views is an alternative, however it is not publicly available for
the time of our query logs, 2006}. The description of our features
follows: 
\begin{itemize}[leftmargin=*]
\item \textbf{Seasonality} is a temporal pattern that indicates how
periodic is an observed behavior over time. We leverage this time series
decomposition technique for detecting not only seasonal events (e.g.,
Christmas Eve, US Open)~\cite{DBLP:conf/sigir/Shokouhi11} but also
more fine-grained periodic ones that recurring on a weekly basis,
such as a TV show program.

\item \textbf{Autocorrelation}, is the cross correlation of a signal with
itself or the correlation between its own past and future values at
different times. We employ autocorrelation for detecting the trending
characteristics of an event, which can be categorized by its
predictability. When an event contains strong inter-day dependencies,
the autocorrelation value will be high. Given observed time series
values ${\psi_{1}, ..., \psi_{N}}$ and its mean $\bar{\psi}$,
autocorrelation is the similarity between observations as a function
of the time lag l between them.  In this work, we consider
autocorrelation at the one time unit lag only (l = 1), which shifts
the second time series by one day.

\item \textbf{Correlation coefficient}, measures the dynamics of two
consecutive aspect ranked lists at time $t_{e}$ and $t_{e}-1$, return
by \textbf{RWR}. We use Goodman and Kruskal's gamma to account for
possible new or old aspects appear or disappear in the newer list.

\item \textbf{Level of surprise}, measured by the error margin in
prediction of the learned model on the time series. This is a good
indicator for detecting the starting time of \textit{breaking}
events. We use Holt-Winters as the predictive model.

\item \textbf{Rising and falling signals.} The intuition behind 
time identification is to measure whether $\mathcal{\psi}^{(e)}_{Q}$
is going up (\textit{before}) or down (\textit{after}) or stays
trending (\textit{during}) at hitting time. Given
$\mathcal{\psi}^{(e)}_{Q}$, we adopt an effective parsimonious model
called SpikeM~\cite{matsubara2012rise}, which is derived from
epidemiology fundamentals to predict the rise and fall of event
diffusion. We use the \textit{Levenberg-Marquardt} algorithm to learn
the parameter set and use the parameters as features for our
classification task.

\end{itemize}

\textit{\textbf{Learning model.}} We assume that there is a 
semantic relation between the event types and times (e.g., the before
phase of \textit{breaking} events are different
from \textit{anticipated}). To leverage the dependency between the
ground labels of the two classification tasks, we apply a joint
learning approach that models the two tasks in a cascaded manner, as a
simple version of ~\cite{heitz2009cascaded}. Given the same input
instance $\mathcal{I}$, the $1^{st}$ stage of the cascaded model
predicts the event type $\mathcal{C}$ with all proposed features. The
trained model $\mathcal{M}^{1}$ is used in the $2^{nd}$ stage to
predict the event time $\mathcal{T}$. We use the logistic regression
model $\mathcal{M}_{LR}^{2}$ for the $2^{nd}$ stage, which allows us
to add additional features from $\mathcal{M}^{1}$. The feature vector
of $\mathcal{M}_{LR}^{2}$ consists of the same features as
$\mathcal{M}^{1}$, together with the probability distribution of
$P(\mathcal{C}_{k}|e,t)$ (output of $\mathcal{M}^{1}$) of as
additional features.

\textit{\textbf{Ranking-sensitive time and type distribution.}} The
output of an effective classifier can be directly used for determining
a time and type probability distribution of entities; and thus
dividing the training entities into subsets for
our \textit{divide-and-conquer} ranking approach. However, having a
pre-learned model with separate and large training data is expensive
and could be detrimental to ranking performance if the training data
is biased. We therefore opt for effective
on-the-fly \textit{ranking-sensitive} time and type identification,
following~\cite{Bian:2010:RSW:1772690.1772705} that utilizes the
`locality property' of feature spaces. We adjust and refine the
approach as follows. Each entity is represented as a feature vector,
and consists of all proposed features with importance weights learned
from a sample of training entities (for ranking). We then employ a
Gaussian mixture model to obtain the centroids of training
entities. In our case, the number of components for clustering are
fixed before hand, as the number of event types multiplied by the
number of event times. Hence the probability distribution of entity
$e$ at time $t$ belonging to time and type
$\mathcal{T}_{l}, \mathcal{C}_{k}$,
$P(\mathcal{T}_{l},\mathcal{C}_{k}|e,t)$ is calculated as $1
- \frac{\mathbf{x}^{e}
- \mathbf{x}_{c_{\mathcal{T}_{l},\mathcal{C}_{k}}}^{2}}{\max_{\forall
T,C} \mathbf{x}^{e} - \mathbf{x}_{c_{\mathcal{T}_{l},\mathcal{C}_{k}}}^{2} }$, or the
distance between feature vector $\mathbf{x}^{e}$ and the corresponding
centroid $c_{\mathcal{T}_{l},\mathcal{C}_{k}}$.

\subsection{Time and Type-Dependent Ranking Models}
\label{subsec:ensemble}
Learning a single model for ranking event entity aspects is not effective due to the dynamic nature of a real-world event driven by a great variety of multiple factors. We address two major factors that are assumed to have the most influence on the dynamics of events at aspect-level, i.e., time and event type. Thus, we propose an adaptive approach based on the ensemble of multiple ranking models learned from training data, which is partitioned by entities' temporal and type aspects. In more detail, we learn multiple models, which are co-trained using data \textit{soft} partitioning / clustering method in Section~\ref{subsec:timeandtopic}, and finally combine the ranking results of different models in an ensemble manner. This approach allows sub-models to learn for different types and times (where feature sets can perform differently), without hurting each other. The adaptive global loss then co-optimizes all sub-models in a unified framework. We describe in details as follows.

\textbf{Ranking Problem.} For aspect ranking context, a typical ranking problem is to find a function $\mathsf{f}$ with a set of parameters $\omega$ that takes aspect suggestion feature vector $\mathcal{X}$ as input and produce a ranking score $\hat{y}$: $\hat{y} = \mathsf{f}(\mathcal{X}, \omega)$. In a learning to rank paradigm, it is aimed at finding the best candidate ranking model $\mathsf{f^{*}}$ by minimizing a given loss function $\mathcal{L}$ calculated as: $\mathsf{f^{*}} = \arg \min_{f}\sum_{\forall a} \mathcal{L}(\hat{y_{a}},y_{a})$.


\textbf{Multiple Ranking Models.} We learn multiple ranking models trained using data constructed from different time periods and types, simultaneously, thus producing a set of ranking models $\mathbf{M} = \left\{M_{\mathcal{T}_1,\mathcal{C}_1}, \ldots, M_{\mathcal{T}_m,\mathcal{C}_n}\right\}$, where $\mathcal{T}_i$ is an event time period, $\in \mathcal{T}$, and $\mathcal{C} = \left\{\mathcal{C}_1,\mathcal{C}_2,\ldots,\mathcal{C}_n\right\}$ are the types of an event entity. We use an ensemble method that combines results from different ranking models, each corresponding to an identified ranking-sensitive query time $\mathcal{T}$ and entity type $\mathcal{C}$. The probabilities that an event entity $e$ belongs to time period $\mathcal{T}_{l}$ and type $\mathcal{C}_{k}$ given the hitting time $t$ is $P(\mathcal{T}_{l},\mathcal{C}_{k}|e,t)$, and can be computed using the time and type identification method presented in Section~\ref{subsec:timeandtopic}.

\vspace{-0.2cm}
\begin{equation} 
\mathsf{f^{*}} = \arg \min_{f}\sum_{\forall a} \mathcal{L}(\sum\limits_{k=1}^{n}P(\mathcal{C}_{k}|a,t)\sum\limits_{l=1}^{m}P(\mathcal{T}_{l}|a,t,\mathcal{C}_{k})\hat{y_{a}},y_{a})
\end{equation}
\vspace{-0.2cm}
%
%

\textbf{Multi-Criteria Learning.} Our task is to minimize the global relevance loss function, which evaluates the overall training error, instead of assuming the independent loss function, that does not consider the correlation and overlap between models. We adapted the L2R RankSVM~\cite{joachims2006training}. The goal of RankSVM is learning a linear model that minimizes the number of discordant pairs in the training data. We modified the objective function of RankSVM following our global loss function, which takes into account the temporal feature specificities of event entities. The temporal and type-dependent ranking model is learned by minimizing the following objective function:

\vspace{-0.2cm}
\begin{equation}
\begin{split}
\min_{\omega,\xi,e,i,j} \frac{1}{2} ||\omega||^{2} + C \sum\limits_{e,i,j} \xi_{e,i,j} \\
 \mbox{subject to,   }  \sum\limits_{k=1}^{n}P(\mathcal{C}_{k}|e,t)\sum\limits_{l=1}^{m}P(\mathcal{T}_{l}|e,t,\mathcal{C}_{k})\omega_{kl}^{T}X_{i}^{e} \\ \geq \sum\limits_{k=1}^{n}P(\mathcal{C}_{k}|e,t)\sum\limits_{l=1}^{m}P(\mathcal{T}_{l}|e,t,\mathcal{C}_{k})\omega_{kl}^{T}X_{j}^{e}
      + 1 - \xi_{e,i,j}, \\ 
      \forall X_{i}^{e} \succ X_{j}^{e}, \xi_{e,i,j} \geq 0.
\end{split}
\end{equation}

where $P(\mathcal{C}_{k}|e,t)$ is the probability the event entity $e$, at time $t$, is of type $\mathcal{C}_{k}$, and $P(\mathcal{T}_{l}|e,t,\mathcal{C}_{k})$ is probability $e$ is in this event time $\mathcal{T}_{l}$ given the hitting-time $t$ and  $\mathcal{C}_{k}$. The other notions are inherited from the traditional model ($X_{i}^{q} \succ X_{j}^{e}$ implies that an entity aspect $i$ is ranked ahead of an aspect $j$ with respect to event entity $e$. $C$ is a trade-off coefficient between the model complexity $||\omega||$ and the training error $\xi_{a,i,j}$.

\textbf{Ensemble Ranking}. After learning all time and type-dependent sub models, we employ
an unsupervised ensemble method to produce the final ranking score. Supposed $\bar{a}$ is a testing entity aspect of entity $e$. We run each of the  ranking models in $\mathbf{M}$ against the instance of $\bar{a}$, multiplied by the time and type probabilities of the associated entity $e$ at hitting time $t$. Finally, we sum all scores produced by all ranking models to obtain the ensemble ranking, $
score(\bar{a}) = \sum_{m \in M} P(\mathcal{C}_{k}|e,t) P(\mathcal{T}_{l}|e,t,\mathcal{C}_{k}) \mathsf{f^{*}}_{m}(\bar{a})$.
	 
\subsection{Ranking Features} 
We propose two sets of features, namely, (1) \textit{salience} features (taking into account the general importance of candidate aspects) that mainly mined from Wikipedia and (2) \textit{short-term interest} features (capturing a trend or timely change) that mined from the query logs. In addition, we also leverage click-flow relatedness features computed using RWR. The features from the two categories are explained in details as follows.

\subsubsection{\textit{Salience} features} - or in principle, \textsf{long-term} prominent features. 
\begin{itemize}[leftmargin=*]
\item \textbf{TF.IDF} of an aspect a is the average $TF.IDF(w)$ of all terms $w \in a$; $TF.IDF(w)$ is calculated as $tf(w,D) \dot log\dfrac{N}{df(w)}$, whereas $D$ is a section in the related Wikipedia articles $C$ of entity $e$. To construct $C$, we take all in-link articles of the corresponding Wikipedia article of $e$; $tf(w,D)$ is the term frequency, $df(w)$ denotes the number of sections which $w$ appears.

\item \textbf{MLE-based}, where we reward the more (cumulated) frequently occurring aspects from the query logs. The maximum likelihood $\mathsf{s}_{MLE}$ is ${\small\dfrac{sum_{w\in a} \mathit{n}(w,e)}{\sum_{a\prime} \sum_{t\in a\prime} \mathit{f}(w,e)}}$, where $\mathit{f}(w,e)$ denotes the frequency a segment (word or phrase) $w \in a$ co-occurs with entity $e$.

\item \textbf{Entropy-based}, where we reward the more ``stable'' aspects over time from the query logs. The entropy is calculated as:
$\mathsf{s}_{E} = \sum_{t\in T} P(a|t,e)logP(a|t,e)$, where $P(a|t,e)$ is the probability of observing aspect $a$ in the context of entity $e$ at time $t$.

\item \textbf{Language Model-based}, how likely aspects are generated by as stastical LM based on the textual representation of the entity $\mathsf{d}(e)$. We model $\mathsf{d}(e)$ as the corresponding Wikipedia article text. We use the unigram model with default Dirichlet smoothing.

\end{itemize}

%



\subsubsection{\textit{Short-term} interest features}, are described as follows.
\begin{itemize}[leftmargin=*]
\item \textbf{Temporal click entropy.} Click entropy~\cite{Dou:2007:LEA:1242572.1242651} is known  as the measurement of how much diversity of clicks to a particular query over time. In detail, the click entropy is measured as the query click variation over a set of URLs for a given query $q$.  In this work, a temporal click entropy accounts for only the number of clicks on the time unit that the entity query is issued. The temporal click entropy $TCE_{t}$ can be computed as $\sum\limits_{u \in U_q} -P(u|q) \log P(u|q)$ where $U_q$ is a set of clicked URLs for a given query $q$ at time $t$. The probability of $u$ being clicked among all the clicks of q, $P(u|q)$ is calculated as $\frac{|\textit{click}(u,q)|}{\sum_{u_i \in U_q}|\textit{click}(u_i,q)|}$.

\item \textbf{Trending momentum} measures the trend of an aspect based on the query volume. The trending momentum at time t, $\mathit{Tm}_{t}$ is calculated using the moving average ($\mathit{Ma}$) technique, i.e., $\mathit{Tm}_{t} = \mathit{Ma}(t,i_{s}) - \mathit{Ma}(t,i_{l})$. Whereas, $i_{s}$,$i_{l}$ denotes the short and long time window from the hitting time.

\item \textbf{Cross correlation} or temporal similarity, is how correlated the aspect \textit{wrt.} the main entity. The more cross-correlated the temporal aspect to the entity, the more influence it brings to the global trend. Given two time series $\mathcal{\psi}^{e}_{t}$ and $\mathcal{\psi}^{a}_{t}$ of the entity and aspect at time t, we employ the cross correlation technique to measure such correlation.  Cross correlation $CCF(\mathcal{\psi}^{e}_{t},\mathcal{\psi}^{a}_{t})$ gives the correlation score at lagging times. Lagging time determines the time delay between two time-series. In our case, as we only interest in the hitting time, we take the maximum $CCF$ in a lag interval of $[-1,1]$.

\item \textbf{Temporal Language Model-based}, similar to the \textit{salient} feature, only the textual representation $d(e)$ is the aggregated content of top-k most clicked URLs at time $t$.

\end{itemize}

\section{Evaluation}
\label{sec:experiments}
In this section, we explain our evaluation for assessing the
performance of our proposed approach. We address three main research
questions as follows:



\textbf{RQ1}: How good is the classification method in identifying the
most relevant event type and period with regards to the hitting time? 

\textbf{RQ2}: How do long-term salience and short-term interest
features perform at different time periods of different event types?


\textbf{RQ3}: How does the ensemble ranking model perform compared to
the single model approaches? 

In the following, we first explain our experimental setting including
the description of our query logs, relevance assessment, methods and
parameters used for the experiments. We then discuss experimental
results for each of the main research questions.

\subsection{Experimental Setting}
\label{subsec:setting}

\textbf{Datasets.}  We use a real-world query log dataset from AOL,
which consists of more than 30 million queries covering the period
from March 1, to May 31, 2006. Inspired by the taxonomy of
event-related queries presented
in~\cite{DBLP:conf/icwsm/KairamMTLD13}, we manually classified the
identified events into two distinct subtypes (i.e., \textit{Breaking}
and \textit{Anticipated}). We use
Tagme~\footnote{\url{https://tagme.d4science.org/tagme/}} to link
queries to the corresponding Wikipedia pages. We use the English
Wikipedia dump of June, 2006 with over 2 million articles to
temporally align with the query logs. The Wikipedia page edits source
is from 2002 up to the studied time, as will be explained later. To
count the number of edits, we measure the difference between
consecutive revision pairs extracted from the
Special:Export~\footnote{\url{https://en.wikipedia.org/wiki/Special:Export}}.

\textit{\textbf{Identifying event entities.}} We reuse the
event-related queryset from~\cite{kanhabua2015learning}, that contains
837 entity-bearing queries. We removed queries that refer to past and
future events and only chose the ones which occured in the period of
the AOL dataset, which results in 300 distinct entity
queries. Additionally, we construct a more recent dataset which
consists of the volume of searches for 500 trending entity queries on
Google Trend. The dataset covers the period from March to May,
2017. To extract these event-related queries, we relied on the
Wikipedia Portal:Current
events\footnote{\url{https://en.wikipedia.org/wiki/Portal:Current_events}}
as the external indicator, as we only access Google query logs via
public APIs. Since the click logs are missing, the Google Trend
queryset is used only as a supplementary dataset for \textit{RQ1}.

\textbf{Dynamic Relevance Assessment.}  There is no standard
ground-truth for this novel task, so we relied on manual annotation to
label entity aspects dynamically; with respect to the studied times
according to each event period. We put a range of 5 days before the
event time as \textit{before} period and analogously for
\textit{after}. We randomly picked a day in the 3 time periods for the
studied times. In our annotation process, we chose 70 popular and
trending event entities focusing on two types of events, i.e.,
\textit{Breaking} (30 queries) and \textit{Anticipated} (40 queries).
For each entity query, we make used of the top-k ranked list of
candidate suggestions generated by RWR,
cf. Section~\ref{subsec:graph}. Four human experts were asked to
evaluate a pair of a given entity and its aspect suggestion (as
relevant or non-relevant) with respect to the event period. We defined
4 levels of relevance: 3 (very relevant), 2 (relevant), 1 (irrelevant)
and 0 (don't know). Finally, 4 assessors evaluated 1,250
entity/suggestion pairs (approximately 3,750 of triples), with
approximately 17 suggestions per trending event on average. The
average Cohen's Kappa for the evaluators' pairwise inter-agreement is
k = 0.78. Examples of event entities and suggestions with dynamic
labels are shown in Table~\ref{tab:dynamiclabel}. The relevance
assessments will be made publicly available.



\begin{table}[t]
  \centering
{\scriptsize
  \caption{Dynamic relevant assessment examples.} 
    \begin{tabular}{llccr}
    \toprule
    \multicolumn{1}{c}{\multirow{2}[0]{*}{Entity}} & \multicolumn{1}{c}{\multirow{2}[0]{*}{Suggestion}} & \multicolumn{3}{c}{Dynamic Label} \\ 
          &       & Before & During & After \\    \midrule
    kentucky derby & + odds & VR & VR & R \\
    kentucky derby & + contenders & VR & R & R \\
    kentucky derby &  + winner & NR & R & VR \\
    kentucky derby & + results & NR & VR & VR \\   
    \bottomrule
    \end{tabular}%
  \label{tab:dynamiclabel}%
  }
  \vspace*{-5pt}
\end{table}%


\textbf{Methods for Comparison.} Our baseline method for aspect
ranking is RWR, as described in Section~\ref{subsec:graph}. Since we
conduct the experiments in a query log context, time-aware query
suggestions and auto-completions (QACs) are obvious competitors. We
adapted features from state-of-the-art work on time-aware QACs as
follows. For the QACs' setting, entity name is given as prior.  Instead of making a direct comparison to the linear models in ~\cite{reinanda2015mining} -- that are tailored to a different variant of our target -- we opt for the supervised-based approach, $SVM_{salient}$, which we consider a fairer and more relevant salient-favored competitor for our research questions. 

\textit{Most popular completion}
(\textbf{MLE})~\cite{Bar-Yossef:2011:CQA:1963405.1963424} is a
standard approach in QAC. The model can be regarded as an approximate
Maximum Likelihood Estimator (MLE), that ranks the suggestions based
on past popularity. Let $P(q)$ be the probability that the next query
is q. Given a prefix $x$, the query candidates that share the prefix
$\mathcal{Q}_{c}$, the most likely suggestion $q \in \mathcal{Q}_{c}$
is calculated as: $MLE(x) = argmax_{q \in \mathcal{Q}_{c}} P(q)$. To
give a fair comparison, we apply this on top of our aspect extraction
cf. Section~\ref{subsec:graph}, denoted as $RWR+MLE$; analogously with
recent MLE.

\textit{Recent MLE}
(\textbf{MLE-W})~\cite{Whiting:2014:RRQ:2566486.2568009,Shokouhi:2012:TQA:2348283.2348364}
does not take into account the whole past query log information like
the original MLE, but uses only recent days. The popularity of query
$q$ in the last $n$ days is aggregated to compute $P(q)$.

\textit{Last N query distribution}
(\textbf{LNQ})~\cite{Whiting:2014:RRQ:2566486.2568009,Shokouhi:2012:TQA:2348283.2348364}
differs from MLE and W-MLE and considers the last $N$ queries given
the prefix $x$ and time $x_{t}$. The approach addresses the weakness
of W-MLE in a time-aware context, having to determine the size of the
sliding window for prefixes with different popularities. In this
approach, only the last $N$ queries are used for ranking, of which $N$
is the trade-off parameter between \textit{robust} (non time-aware
bias) and \textit{recency}.

\textit{Predicted next N query distribution} (\textbf{PNQ}) employs
the past query popularity as a prior for predicting the query
popularity at hitting time, to use this prediction for
QAC~\cite{Whiting:2014:RRQ:2566486.2568009,Shokouhi:2012:TQA:2348283.2348364}. We
adopt the prediction method proposed
in~\cite{Shokouhi:2012:TQA:2348283.2348364}.

\textbf{Parameters and settings.} The jumping probability for RWR is
set to 0.15 (default). For the classification task, we use models
implemented in Scikit-learn~\footnote{\url{http://scikit-learn.org/}}
with default parameters. For learning to rank entity aspects, we
modify RankSVM.  For each query, the hitting time is the same as used
for relevance assessment.  Parameters for RankSVM are tuned via grid
search using 5-fold cross validation (CV) on training data, trade-off
$c =20$. For W-MLE, we empirically found the sliding window $W = 10$
days. The time series prediction method used for the PNQ baseline and
the prediction error is Holt-Winter, available in R. In LNQ and PNQ,
the trade-off parameter N is tuned to 200. The short-time window
$i_{s}$ for the trending momentum feature is 1-day and long $i_{l}$ is
5-days. Top-k in the temporal LM is set to 3. The time granularity for
all settings including hitting time and the time series binning is 1
day.

For RQ1, we report the performance on the \textit{rolling} 4-fold CV
on the whole dataset. To seperate this with the L2R settings, we
explain the evaluating methodology in more details in
Section~\ref{subsec:cc}. For the ranking on partitioned data (RQ2), we
split \textit{breaking} and \textit{anticipated} dataset into 6
sequential folds, and use the last 4 folds for testing in a rolling
manner. To evaluate the ensemble method (RQ3), we use the first two
months of AOL for training (50 queries, 150 studied points) and the
last month (20 queries as shown in Table~\ref{tab:20test}, 60 studied
points) for testing.

\begin{table}[t]
\centering
\caption{Example entities in May 2006.}
\label{tab:20test}
{\scriptsize
\begin{tabular}{|l|l|}
\hline
\multirow{3}{*}{\textbf{anticipated}} & may day, da vinci code, cinco de mayo, american idol,                     \\ \cline{2-2} 
                                      & anna nicole smith, mother's day, danica patrick, emmy rossum,             \\ \cline{2-2} 
                                      & triple crown, preakness stakes, belmont stakes kentucky derby, acm awards \\ \hline
\multirow{2}{*}{\textbf{breaking}}    & david blaine, drudge report, halo 3, typhoon chanchu,                     \\ \cline{2-2} 
                                      & patrick kennedy, indonesia,  heather locklear                             \\ \hline
\end{tabular}
}
\end{table}

\textbf{Metrics.}  For assessing the performance of classification
methods, we measured accuracy and F1. For the retrieval effectiveness
of query ranking models, we used two metrics, i.e., Normalized
Discounted Cumulative Gain (NDCG) and $recall@k$ ($r@k$). We measure the retrieval effectiveness of each metric at 3 and 10 ($m$@3 and $m$@10, where $m \in$ $\left\{NDCG,
R\right\}$). $NDCG$ measures the ranking
performance, while $recall@k$ measures the proportion of relevant
aspects that are retrieved in the top-k results.
\vspace{-0.2cm}
\vspace{-0.2cm}
\subsection{Cascaded Classification Evaluation}
\label{subsec:cc}
\textbf{Evaluating methodology.} 
For \textbf{RQ1}, given an event entity e, at time t, we need to classify them into either \textit{Breaking} or \textit{Anticipated} class. We select a studied time for each event period randomly in the range of 5 days before and after the event time. In total, our training dataset for AOL consists of 1,740 instances of \textit{breaking} class and  3,050 instances of \textit{anticipated}, with over 300 event entities. For \textit{GoogleTrends}, there are 2,700 and 4,200 instances respectively. We then bin the entities in the two datasets chronologically into 10 different parts. We set up 4 trials with each of the last 4 bins (using the history bins for training in a \textit{rolling} basic) for testing; and report the results as average of the trials. 

\textbf{Results.} The baseline and the best results of our $1^{st}$ stage event-type classification is shown in Table~\ref{tab:type}-\textbf{top}. The accuracy for basic majority vote is high for imbalanced classes, yet it is lower at weighted F1. Our learned model achieves marginally better result at F1 metric.

\begin{table}[t]
\centering
{\scriptsize
\caption{Event type and time classification performance.}
\label{tab:type}
\begin{tabular}{|l|l|l|l|l|}
\hline
                            & \textbf{Dataset}              & \textbf{Model}      & \textbf{Accuracy} & \textbf{Weighted F1} \\ \hline
\multirow{4}{*}{Event-type} & \multirow{2}{*}{AOL}          & \textbf{majority votes}      & 0.64            & 0.58               \\ \cline{3-5} 
                            &                               & \textbf{SVM                } & \textbf{0.79}   & \textbf{0.89}      \\ \cline{2-5} 
                            & \multirow{2}{*}{GoogleTrends} & \textbf{majority votes}     & 0.61            & 0.68             \\ \cline{3-5} 
                            &                               & \textbf{SVM                } & \textbf{0.83}   & \textbf{0.85}      \\ \hline \hline
\multirow{4}{*}{Event-time} & \multirow{2}{*}{AOL}          & \textbf{Logistic Regression} & 0.68           & 0.72               \\ \cline{3-5} 
                            &                               & \textbf{\textit{Cascaded}}   & \textbf{0.73}   & \textbf{0.83}      \\ \cline{2-5} 
                            & \multirow{2}{*}{GoogleTrends} & \textbf{Logistic Regression} & 0.71            & 0.78              \\ \cline{3-5} 
                            &                               & \textbf{\textit{Cascaded}}   & \textbf{0.75}   & \textbf{0.82}      \\ \hline
\end{tabular}
}
\end{table}

We further investigate the identification of event time, that is learned on top of the event-type classification. For the gold labels, we gather from the studied times with regards to the event times that is previously mentioned. We compare the result of the cascaded model with non-cascaded logistic regression. The results are shown in Table~\ref{tab:type}-\textbf{bottom}, showing that our cascaded model, with features inherited from the performance of SVM in previous task, substantially improves the single model. However, the overall modest results show the difficulty of this multi-class classification task.

\begin{figure*}[t]
\centering
   \begin{subfigure}{0.95\textwidth}
   \includegraphics[width=1\linewidth]{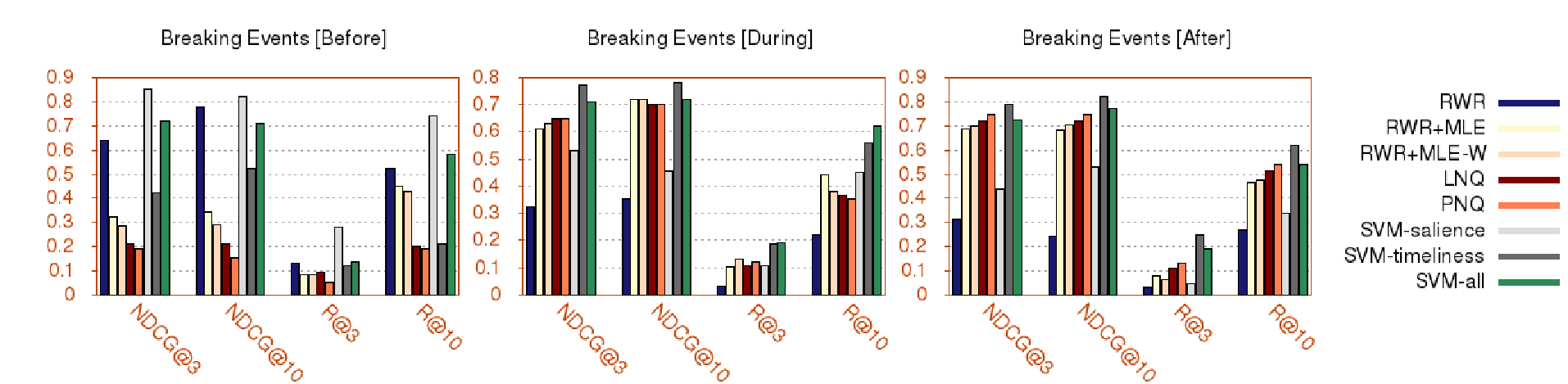}
   \label{fig:Ng1} 
\end{subfigure}

\begin{subfigure}{0.95\textwidth}
   \includegraphics[width=1\linewidth]{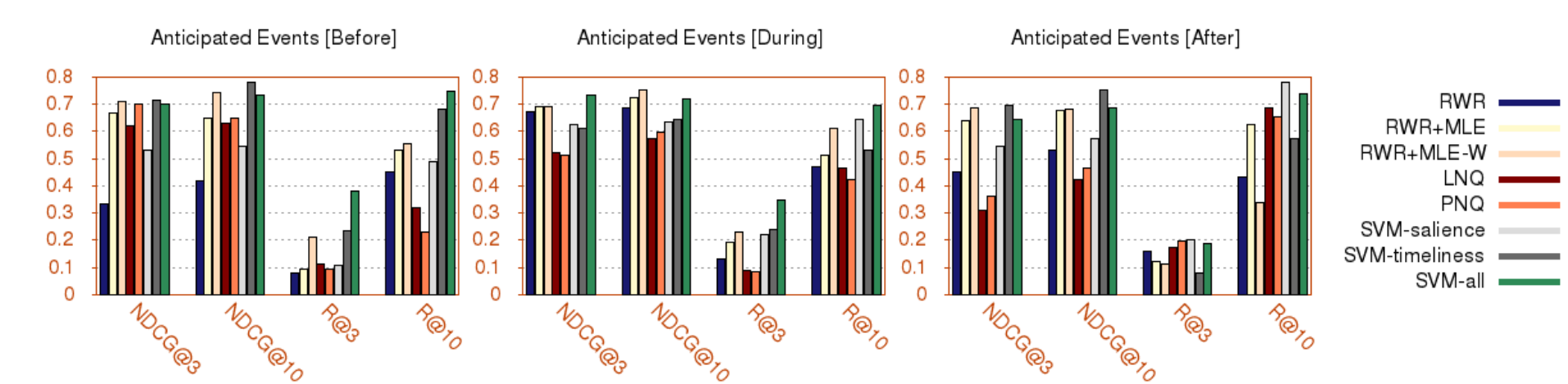}
   \label{fig:Ng2}
\end{subfigure}

\vspace*{-5pt}
\caption{Performance of different models for event entities of different types.}
\label{fig:singles}
\end{figure*}

\vspace*{-5pt}
\subsection{Ranking Aspect Suggestions} 
For this part, we first focus on evaluating the performance of single L2R models that are learned from the \textsf{pre-selected} time (before, during and after) and types (\textit{Breaking} and \textit{Anticipate}) set of entity-bearing queries. This allows us to evaluate the feature performance i.e., \textit{salience} and \textit{timeliness}, with time and type specification (RQ2). We then evaluate our ensemble ranking model (results from the cascaded evaluation) and show it robustly improves the baselines for all studied cases (RQ3). Notice that, we do not use the learned classifier in Section~\ref{subsec:cc} for our ensemble model, since they both use the same time period for training, but opt for the \textit{on-the-fly} ranking-sensitive clustering technique, described in Section~\ref{subsec:timeandtopic}.
 

\noindent\textbf{RQ2.} Figure~\ref{fig:singles} shows the performance of the aspect ranking models for our event entities at specific times and types. The most right three models in each metric are the models proposed in this work. The overall results show that, the performances of these models, even better than the baselines (for at least one of the three), vary greatly among the cases. In general, $SVM_{salience}$ performs well at the \textbf{before} stage of breaking events, and badly at the \textbf{after} stage of the same event type. Whereas $SVM_{timeliness}$ gives a contradictory performance for the cases. For anticipated events, $SVM_{timeliness}$ performs well at the \textbf{before} and \textbf{after} stages, but gives a rather low performance at the \textbf{during} stage. For this event type, $SVM_{salience}$ generally performs worse than $SVM_{timeliness}$. Overall, The $SVM_{all}$ with all features combined gives a good and stable performance, but for most cases, \textsf{are not better} than the well-performed single set of features L2R model. In general, these results prove our assumption that \textit{salience} and \textit{timeliness} should be traded-off for different event types, at different event times. For feature importances, we observe regularly, stable performances of \textit{same-group} features across these cases. \textit{Salience} features from knowledge bases tend to perform better than from query logs for \textit{short-duration} or less popular events. We leave the more in-depth analysis of this part for future work.

\noindent\textbf{RQ3.} We demonstrate the results of single models and our ensemble model in Table~\ref{tab:knn_all_fix}. As also witnessed in RQ2,  $SVM_{all}$, will all features, gives a rather stable performance for both NDCG and Recall, improved the baseline, yet not significantly. Our \textit{Ensemble} model, that is learned to trade-off between \textit{salience} and \textit{timeliness} achieves the best results for all metrics, outperforms the baseline significantly.  As the testing entity queries in this experiment are at all event times and with all event types, these improvements illustrate the robustness of our model.  Overall, we witness the low performance of adapted QAC methods. One reason is as mentioned, QACs, even time-aware generally favor already \textit{salient} queries as follows the \textit{rich-get-richer} phenomenon, and are not ideal for entity queries that are event-related (where aspect relevance can change abruptly). Time-aware QACs for partially long prefixes like entities often encounter sparse traffic of query volumes, that also contributes to the low results.
\vspace{-0.cm}
\begin{table}[t]
  \centering
{\scriptsize
  \caption{Performance of the baselines (RWR relatedness scores, RWR+MLE, RWR+MLE-W, LNQ, and PNQ) compared with our ranking models;
$\ast$,$\dagger$, $\mp$ indicates statistical improvement over the baseline using t-test with significant at $p<0.1$, $p<0.05$, $p<0.01$ respectively.}
    \begin{tabular}{clrrrr}
        \toprule
              & Methods &  NDCG@3 & NDCG@10 & R@3 & R@10 \\
        \midrule
      
        \multicolumn{1}{c}{} & RWR & \textit{0.3208} & \textit{0.4137} & \textit{0.1208} & \textit{0.3749} \\ 
        \multicolumn{1}{c}{} & RWR+MLE   & +29.94\% & +9.73\% & -21.09\% & +5.15\%$\ast$ \\
        \multicolumn{1}{c}{} & RWR+MLE-W & +11.56\% & +11.46\% & -18.93\%$\ast$ & +3.28\% \\
        \multicolumn{1}{c}{} & LNQ   & +15.39\% & -3.75\% & -19.74\% & -30.31\% \\
        \multicolumn{1}{c}{} & PNQ   & +13.19\% & -9.95\% & -23.46\% & -33.53\% \\ \cmidrule{2-6}
        \multicolumn{1}{c}{} & $SVM_{salience}$ & +41.75\%$\ast$ & +9.18\% & +23.32\%$\ast$ & +9.93\% \\
        \multicolumn{1}{c}{} & $SVM_{timeliness}$ & +15.19\% & +17.53\% & +14.77\% & +11.3\% \\
        \multicolumn{1}{c}{} & $SVM_{all}$ & +52.65\%$\ast$ & +40.87\%$\ast$ & +9.73\%$\dagger$ & +24.3\% \\ \cmidrule{2-6}
				\multicolumn{1}{c}{} & \textbf{Ensemble} & \textbf{+85.12}\%$\mp$ & \textbf{+45.34}\%$\dagger$ & \textbf{+42.78}\%$\ast$ & \textbf{+17.45}\%$\ast$ \\ \midrule
        \bottomrule
    \end{tabular}%
  \label{tab:knn_all_fix}%
  }
\end{table}

\vspace{-0.2cm}	
\section{Conclusion}
\label{sec:conclusion}

We studied the temporal aspect suggestion problem for entities in
knowledge bases with the aid of real-world query logs. For each
entity, we ranked its temporal aspects using our proposed novel time
and type-specific ranking method that learns multiple ranking models
for different time periods and event types. Through extensive
evaluation, we also illustrated that our aspect suggestion approach
significantly improves the ranking effectiveness compared to
competitive baselines. In this work, we focused on a ``global''
recommendation based on public attention. The problem is also
interesting taking other factors (e.g., \textit{search context}) into
account, which will be interesting to investigate in future work.

\def\bibfont{\scriptsize}
\bibliographystyle{abbrv}
{\scriptsize
\bibliography{newbib}
}
\end{document}